\documentclass{article}
\usepackage{float}
\usepackage[utf8]{inputenc}
\usepackage[caption = false]{subfig}
\usepackage{spconf,amsmath,graphicx}
\usepackage{xcolor}
\usepackage{hyperref}
\usepackage{lipsum,multicol}

\usepackage{amsfonts}
\usepackage{ulem}
\usepackage{tikz}
\usepackage{stmaryrd}
\usepackage{amsthm}
\usepackage{enumitem}
\usepackage{multirow}

\usetikzlibrary{automata, arrows.meta, positioning}

\usepackage{enumitem}
\graphicspath{ {img/} }
\title{Combining Progressive Image Compression and Random Access in DNA Data Storage\thanks{This work was funded by the European Union's Horizon research and innovation programme projects Glaciation (Grant No. 101070141), and ANR PEPR program MoleculArxiv\\ 979-8-3315-1213-2/25/\$31.00 \copyright 2025 IEEE}}
\name{Xavier Pic$^{\star}$
, Raja Appuswamy$^{\star}$}
\address{$^{\star}$ EURECOM, Data Science Department, Sophia Antipolis, France}

\begin{document}
\maketitle
\begin{abstract}
\vspace*{-0.5\baselineskip}
The exponential increase in storage demand and low lifespan of data storage devices has resulted in long-term archival and preservation emerging as a critical bottlenecks in data storage. 
In order to meet this demand, researchers are now investigating novel forms of data storage media.
The high density, long lifespan and low energy needs of synthetic DNA make it a promising candidate for long-term data archival. However, current DNA data storage technologies are facing
challenges with respect to cost (writing data to DNA is expensive) and reliability (reading and writing data is error prone). Thus, data compression and error correction are crucial to scale DNA storage.
Additionally, the DNA molecules encoding several files are very often stored in the same place, called an oligo pool. For this reason, without random access solutions, it is relatively impractical to decode a specific file from the pool, because all the oligos from all the files need to first be sequenced, which greatly deteriorates the read cost. 

This paper introduces PIC-DNA--a novel JPEG2000-based progressive image coder adapted to DNA data storage. This coder directly includes a random access process in its coding system, allowing for the retrieval of a specific image from a pool of oligos encoding several images. The progressive decoder can 
dynamically adapt the read cost according to the user's cost and quality constraints at decoding time. Both the random access and progressive decoding greatly improve on the read-cost of image coders adapted to DNA.

\end{abstract}
\vspace*{-0.25\baselineskip}
\begin{keywords}
DNA data storage, JPEG 2000, progressive, random access, JPEG DNA VM
\end{keywords}
\vspace*{-1\baselineskip}
\section{Introduction}
\vspace*{-\baselineskip}
The demand for data storage, fueled by an ever increasing use of social media and video streaming services, poses unprecedented challenges for data storage providers, as they have to store increasingly large-amounts of data over very long periods of time while maintaining retrieval guaranties. 
Conventional storage devices, unfortunately, have fundamental durability and density limitations that make long-term data storage infeasible.
As a result, novel data storage solutions, such as synthetic DNA molecules, have emerged as an alternative for overcoming these challenges \cite{Church},
mainly due to their long lifespan, high density and low energy needs (a DNA molecule can be safely stored in hermetic capsules without consuming any electricity). 

A classic solution for storing data onto DNA molecules is composed of both biochemical and computational processes. The two main biochemical processes are synthesis, to create DNA molecules with the desired sequences of nucleotides, and sequencing, to retrieve the sequence of nucleotides that it contains. The computational processes necessary to store data onto DNA is a DNA-adapted codec, where an encoder transforms digital information into quaternary sequences that are transformed into sequences of nucleotides by synthesis, and a decoder does the reverse job of converting sequencing-generated quaternary sequences (also called reads) back into images. The high costs of the biochemical processes (both in terms of time and money) mean that it is crucial to encode the data into quaternary codes that are as compressed as possible. 

Modern data storage systems that are used for housing large image collections also support efficient random access that enables retrieval of a subset of images. Such random access is used extensively by modern-day web applications to support novel features like adaptive selection of image resolution based on the client (desktop or mobile) used to view the image. Unfortunately, DNA, by itself, does not support such random access mechanisms. Prior work on DNA data storage has demonstrated the use of Polymerase Chain Reaction (PCR) for enabling random access. PCR requires the use of primers, which are short sequences of nucleotides attached to an DNA that can be used to biochemically select, copy and sequence only the oligos that include this specific primer pair. By attaching different primers with different oligos, prior work has demonstrated the ability to achieve random access on DNA. However, such random access has been applied to arbitrary binary files and has not been used into the context of image storage for features like adaptive selection of resolution based on progressive image coding.

In this paper, we present a Progressive Image compression for DNA storage (PIC-DNA)--a novel codec that integrates random access at the image coding level to enable new access paths over image collections stored in DNA and reduce the reading cost. Our coding solution is based on the JPEG 2000 \cite{JPEG2000} progressive coder, but can easily be adapted to any progressive coder that separates the encoded bitstream into resolution or quality layers. PIC-DNA encodes and stores each quality layer separately, using primers to enable selective amplification of specific layers.
By doing so, PIC-DNA enables several access modalities. First, it enables the use of  thumbnails encoded in DNA as an interface between the user and the oligo memory; through the visual selection of the thumbnail, the user can select the image to be decoded, and with the help of the oligos encoding the selected thumbnail, the rest of the image data can be retrieved. Second, it enables each image retrieval from DNA storage to be customized based on the desired resolution. For instance, applications that need only a low quality version of the image can use primers tagged for lower resolution to selectively amplify only a few layers. This results in a lower cost of sequencing, as fewer oligos need to be sequenced, and lower computational complexity of decoding, as fewer reads need to be processed.
To our knowledge, the solution presented in this paper is the first DNA image coder that directly includes a random access mechanism.


\vspace*{-1.25\baselineskip}
\section{Context}
\vspace*{-\baselineskip}
\subsection{DNA-adapted coding}
\vspace*{-0.5\baselineskip}
The field of DNA data storage has emerged as a very active research field over the past decade. In this section, we provide an overview of a few pioneering approaches and refer the reader to recent surveys~\cite{dna-storage-survey} for a detailed comparison of various approaches that focus on storing generic, binary data using DNA. In 2012, Church et al. \cite{Church}, introduced an approach to enable large-scale encoding and decoding of data into synthetic DNA molecules. This work also identified some of the constraints that need to be respected when coding data for synthesized DNA molecules.
Later, in 2013, Goldman et al.\cite{Goldman} provided one of the first encoders capable of respecting some of these constraints. The contribution of that work, an entropy coder, allowed for the encoding of any file into DNA-like data. In 2015, Grass et al.\cite{Grass} introduced the the first error correction codes into a DNA data storage solution. This error correction mechanism makes the whole storage process robust against the different biochemical operations that often introduce errors (substitutions, insertions, deletions) in the encoded DNA-like data. Following this, other works introduced other error correction solutions \cite{Erlich,Yazdi,Aeon,Blawat} with the focus of adding redundancy to binary data to detect and correct errors. Finally, random access solutions aiming to enable selective access to binary data stored in DNA and improve read cost have also been investigated \cite{organick}.
\vspace*{-1.25\baselineskip}
\subsection{DNA-adapted image coding}
\vspace*{-0.5\baselineskip}
While the solutions described above focused on generic binary data storage, several solutions have also emerged to customize the encoding and storage of images specifically on DNA. For instance, Dimopoulou et al.\cite{Dimopoulou} developed a JPEG-based image coder adapted to DNA data storage that uses the Goldman encoder to encode runcat values into DNA.  In \cite{SFC4}, an improved DNA-adapted entropy coder was used, to increase the performance of this JPEG-based DNA-adapted image coder. Lazzarotto et al.\cite{EPFL-Raptor} developed the JPEG DNA VM software that encodes data with a system based on Raptor codes\cite{Raptor-codes}. It was output by the JPEG DNA ad-hoc group as a Verification Model for later developments.

\vspace*{-1\baselineskip}
\section{Proposed Method}
\vspace*{-1\baselineskip}
In this section, we introduce the PIC-DNA codec which adapts progressive image compression to DNA data storage. 
\begin{figure*}[t]
    \centering
    \includegraphics[width=0.6\linewidth]{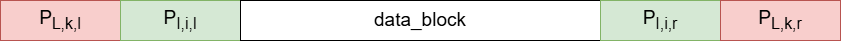}
\vspace*{-0.5\baselineskip}
    \caption{{Organization of an oligo in our solution : the pair of primers $(P_{L,k,l}, P_{L,k,r})$ describes the resolution level to which the oligo belongs, and the pair of primers $(P_{I,i,l}, P_{I,i,r})$ describes the image the oligo is related to.}}
\vspace*{-1\baselineskip}
    \label{fig:oligo_org}
\end{figure*}

\vspace*{-1.25\baselineskip}
\subsection{Coding solution and oligo organization}
\begin{figure*}[h]
\begin{minipage}{.2\textheight}
\hspace*{-0.6cm}
    \subfloat[Progressive Decoding]{\includegraphics[height=0.075\textheight]{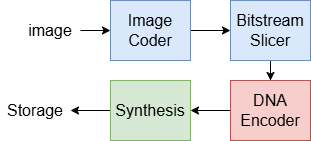}\label{fig:encoding_process}}
    \hspace*{0.2cm}
    \subfloat[Thumbnails extraction]{\includegraphics[height=0.075\textheight]{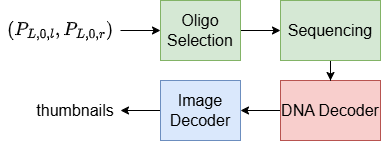}\label{fig:thumbnail_extraction}}
    \hspace*{0.2cm}
    \subfloat[Image primer identification]{\includegraphics[height=0.075\textheight]{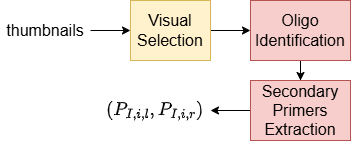}\label{fig:secondary_primer}}
    \hspace*{0.2cm}
    \subfloat[Decoding process]{\includegraphics[height=0.075\textheight]{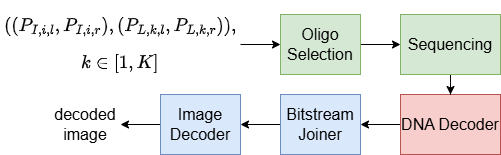}\label{fig:decoding_process}}
    \hspace*{0.2cm}
\end{minipage}
\vspace*{-0.5\baselineskip}
    \caption{ Different components of the general Progressive encoding and decoding workflow}\label{fig:sub}
\vspace*{-1\baselineskip}
\end{figure*}

\vspace*{-0.75\baselineskip}
\noindent\textbf{Data compression.} The general encoding process, as described in Figure \ref{fig:encoding_process}, relies on the binary progressive encoding process of JPEG2000. The image is encoded, with a fixed number of resolution layers $N_{levels}$. The layer with the smallest rate is used as thumbnail. The bitstream is then cut into several binary files, one for each resolution layer.
Each separate layer file is then encoded into DNA with the help of a DNA coder. In our case, we use the transcoder mode of the JPEG DNA VM. 

\noindent\textbf{Random access}: When data is stored into DNA, several files are stored in the same pool. Reading all the files is very costly, so directly accessing the oligos specific to the desired file is crucial. Further, in order to enable progressive decoding, it is also necessary to be able to access each layer separately. In order to enable such access, PIC-DNA two pairs of primers are then concatenated at each end of each oligo as shown in Figure \ref{fig:oligo_org}. The first pair of image primers $(P_{I,i,l}, P_{I,i,r}), i\in\llbracket0,N_{images}-1\rrbracket$ are used to identify each image uniquely. The second pair of layer primers $(P_{L,k,l}, P_{L,k,r}), k\in\llbracket0,N_{levels}-1\rrbracket$ is used to identify each layer uniquely. 

Further, in order enable thumbnail-based image search functionality mentioned earlier, PIC-DNA encodes the thumbnails of all images in the collection separately. These thumbnail oligos follow a similar structure to other data oligos as shown in Figure~\ref{fig:oligo_org}. However, PIC-DNA uses a specific pair of primers $(P_{L,0,l}, P_{L,0,r})$ in place of layer primes to identify the oligos as thumbnail oligos. Second, PIC-DNA uses the same image primer pair for image oligos and thumbnail oligos corresponding to the same image. Thus, in the image primer positions $(P_{I,i,l}, P_{I,i,r}), i\in\llbracket0,N_{images}-1\rrbracket$ of the thumbnail oligos, PIC-DNA stores the primers corresponding to the thumbnail's parent image.

\vspace*{-1.25\baselineskip}
\subsection{Decoding process}
\vspace*{-0.5\baselineskip}
The whole decoding process introduced with our solution can be decomposed into three separate subprocesses: (i) thumbnail extraction, (ii) secondary primer identification, and (iii) decoding.
\begin{enumerate}[leftmargin=*]
\vspace*{-0.5\baselineskip}
\item \textbf{Thumbnails extraction} (Fig. \ref{fig:thumbnail_extraction}): With the help of the pair of primers, $(P_{L,0,l}, P_{L,0,r})$ associated to all the thumbnails, the oligos related to all the thumbnails in the pool are selected. They are then classified image by image, with the help of the secondary image primers, used here as an offset code. The DNA decoder then decodes all the thumbnail oligos into compressed binary files, that are then decoded with the image codec.
    
    
\vspace*{-0.5\baselineskip}
\item \textbf{Secondary primers identification} (Fig. \ref{fig:secondary_primer}): A visual inspection of all the thumbnails allows for the selection of the desired image. Once this image has been identified, the related thumbnail oligos are inspected, and the secondary image primers $(P_{I,i,l}, P_{I,i,r})$ related to the image (as presented in Figure \ref{fig:oligo_org}) can be extracted.
    
    
\vspace*{-0.5\baselineskip}
\item \textbf{Image decoding} (Fig. \ref{fig:decoding_process}):
To decode an image at a certain quality or resolution, the user has to decode all the layers until the chosen resolution layer is reached before they can stop decoding. With PIC-DNA, this has to be done in several stages. First, the user has to run a PCR round to select and augment  oligos related to the desired image. Then multiple rounds of PCR can be performed using this amplified pool to isolate further and augment the oligos of each layer until the target resolution is reached. 
More specifically, once the primers corresponding to the image have been identified from the previous stage, with the help of these primers $(P_{I,i,l}, P_{I,i,r})$ and the pairs of primers related to the different levels to be decoded $\{(P_{L,k,l}, P_{L,k,r}), k\in\llbracket0,K_{levels\_to\_decode}-1\rrbracket\}$, the necessary oligos for decoding the image can now be retrieved. It is important to note that this decoding process, like what is being done in classic binary progressive decoders, can be done iteratively, until a satisfactory version of the image is retrieved. The progressive decoder provides to the end user a trade off between the read-cost (closely linked to the sequencing price and duration), and the quality of reconstruction of the original image. Once the image is decoded, this trade-off can even be re-evaluated, to decode additional resolution layers, since the layers are organized hierarchically. 
\end{enumerate}
\vspace*{-1.75\baselineskip}
\subsection{Read Cost}
\vspace*{-0.5\baselineskip}

Under these conditions, the general reading cost necessary to retrieve a specific image $I$ until the resolution level $K$ can be evaluated in three ways, without random access or progressive decoding (Eq. \ref{eq:def1}), with just progressive decoding (Eq. \ref{eq:def2}), with both progressive decoding and random access (Eq. \ref{eq:def3}):
\vspace*{-0.5\baselineskip}
\begin{equation}
    R_c(I, K) = \frac{\sum_{i=0}^{N_{images}}\sum_{k=0}^{N_{levels}}nucs(i, k)}{input\_image\_pixels}
\label{eq:def1}
\end{equation}
\vspace*{-0.75\baselineskip}
\begin{equation}
    R_{c\_pd}(I, K) = \frac{\sum_{i=0}^{N_{images}}\sum_{k=0}^{K}nucs(i, k)}{input\_image\_pixels}
\label{eq:def2}
\end{equation}
\vspace*{-0.75\baselineskip}
\begin{equation}
    R_{c\_ra}(I, K) = \frac{\sum_{i=0}^{N_{images}}nucs(i, 0) + \sum_{k=1}^{K}nucs(I,k)}{input\_image\_pixels}
\label{eq:def3}
\end{equation}
    The $nucs(i,k)$ value represents the number of nucleotides to sequence to be able to decode a layer:
\vspace*{-0.5\baselineskip}
\begin{equation}
    nucs(i,k) = coverage(i,k) \times number\_oligos(i,k)
\vspace*{-0.5\baselineskip}    
\end{equation}
With this, we can define a read-cost gain as in two different ways, depending on whether or not Random Access is enabled (in both cases, Progressive decoding is used) as:
\vspace*{-0.5\baselineskip}
\begin{equation}
    G_{pd}(I, K) = \frac{R_c(I, K)}{R_{c\_pd}(I, K)}\\
\vspace*{-0.5\baselineskip}
\end{equation}
\begin{equation}
    G_{ra}(I, K) = \frac{R_c(I, K)}{R_{c\_ra}(I, K)}\\
\vspace*{-0.5\baselineskip}
\end{equation}
\vspace*{-1.25\baselineskip}

\section{Experimental Results}
\vspace*{-0.75\baselineskip}
\subsection{Software implementation details}
\vspace*{-0.5\baselineskip}
Our PIC-DNA implementation is currently utilizing the OpenJPEG implementation of the JPEG2000 codec as an image coder. It utilizes the progressive coding functionality that is available in it. In the version 1.5 of the OpenJPEG software, the \texttt{image\_to\_j2k} executable can be configured to output a \texttt{.Idx} file, that gives all the information necessary to localize the different resolution layers in the general output bitstream. Thanks to this, the bitstream is cut at the beginning of every new layer. Two options are available to separate the image into layers: quality layers and resolution layers. Depending on which one is selected, the progression order has to be configured either in RLCP or LRCP encoding (\texttt{-o} option in the \texttt{image\_to\_j2k} software).
\vspace*{-.75\baselineskip}
\subsection{Performance evaluation}
\vspace*{-0.5\baselineskip}
The performance of any DNA encoding method can be evaluated with respect to a series of metrics such as RD-curves, reading cost and writing cost. As our work primarily focuses on progressive compression and random access, the main metric we focus on in our study is the reading cost. We especially study the evolution of the reading cost for a given encoded image, as we read through each resolution layer. The image is encoded into a series of resolutions layers, the smallest one being the thumbnail. The resolution layers each divide the size of the image by a factor of 2 in each dimension (4 in total).

The results presented here were obtained by encoding the images of the kodak\footnote{\url{https://r0k.us/graphics/kodak/}} dataset, with 5 resolution levels. The oligos had data blocks of length 200 to which the primers were added. A theoretical read-cost gain was observed (lines 3 and 7 of Table \ref{tab:read_costs}), for a given layer, setting the coverage to 1 in the gain metrics previously defined.
A noise simulator was used to introduce substitutions, deletions and insertions. Erroneous data was then passed through a clustering and consensus process to estimate a minimal coverage for decodability. 
The results in Table \ref{tab:read_costs} depict the read cost gains obtained by enabling progressive decoding and random access. 

The top part of Table \ref{tab:read_costs} represents the gains obtained when only progressive decoding is enabled. In this situation, we decode all images until a certain resolution level, without any random access. In these conditions, the progressive decoder provides gains of up to 60$\times$, when only the initial layer is targeted. This gain quickly decreases if more layers are targeted, and if all the layers are targeted, no gain can be leveraged from progressive decoding, because all oligos need to be sequenced. It is important to note that here, the gains depend on the dimensions of the chosen resolution layers: smaller thumbnails or layers will leverage better gains, at the cost of more distorted images.

The bottom part of Table \ref{tab:read_costs} represents the gains obtained when both progressive decoding and random access are enabled. In these gains, the cost necessary to retrieve the thumbnail is included, as described in the second member of the sum in Equation (\ref{eq:def3}). The random access process further improves the read cost, especially in layers with better resolution where the gain is multiple times larger than the one measured with only progressive decoding enabled. This gain highly depends on the number of images encoded in the pool of DNA molecules, and on the size of the different resolution layers.

\begin{figure}
\begin{minipage}{.4\textwidth}
    \subfloat[Progressive Decoding]{\includegraphics[width=0.6\textwidth]{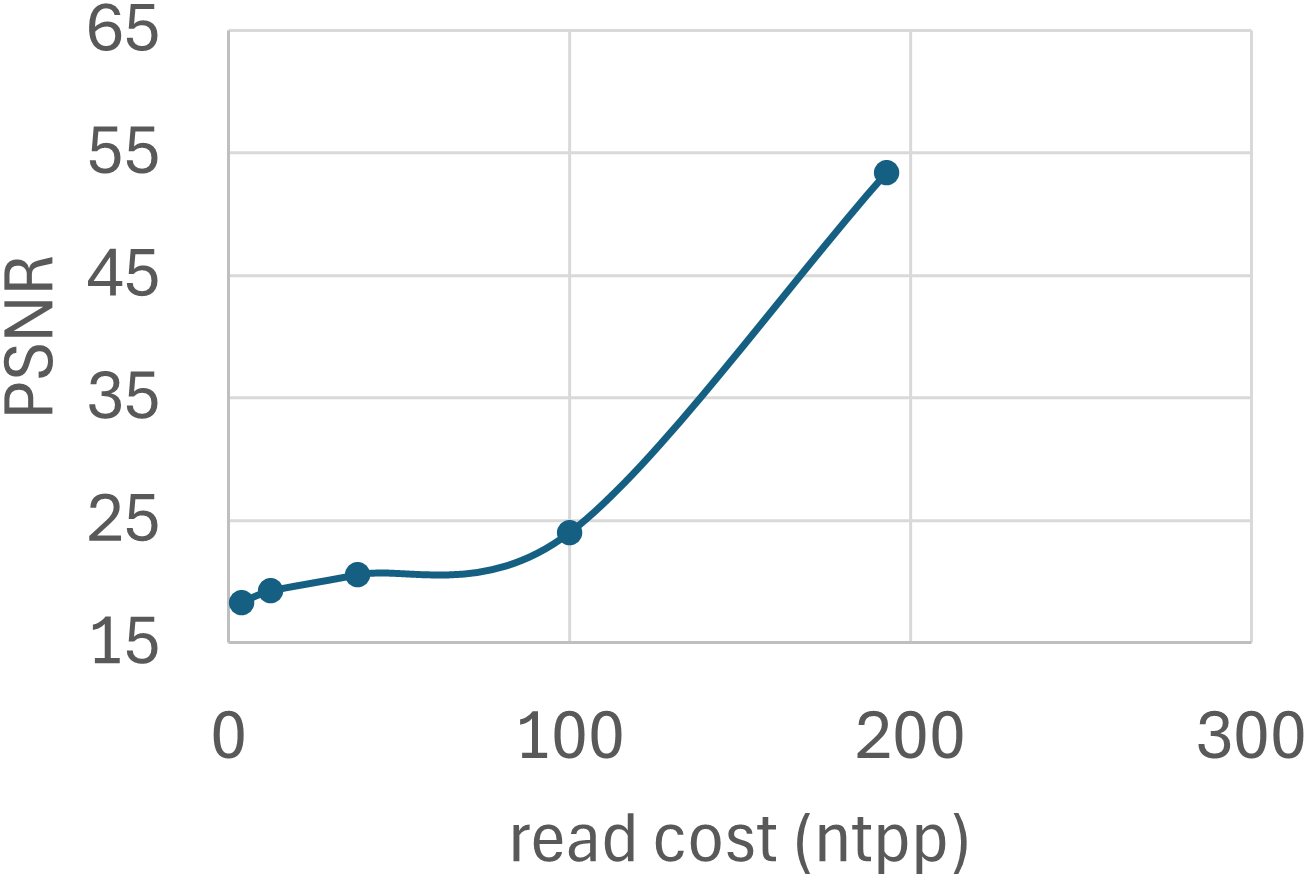}\label{fig:sub_a}}
    \subfloat[Random Access]{\includegraphics[width=0.6\textwidth]{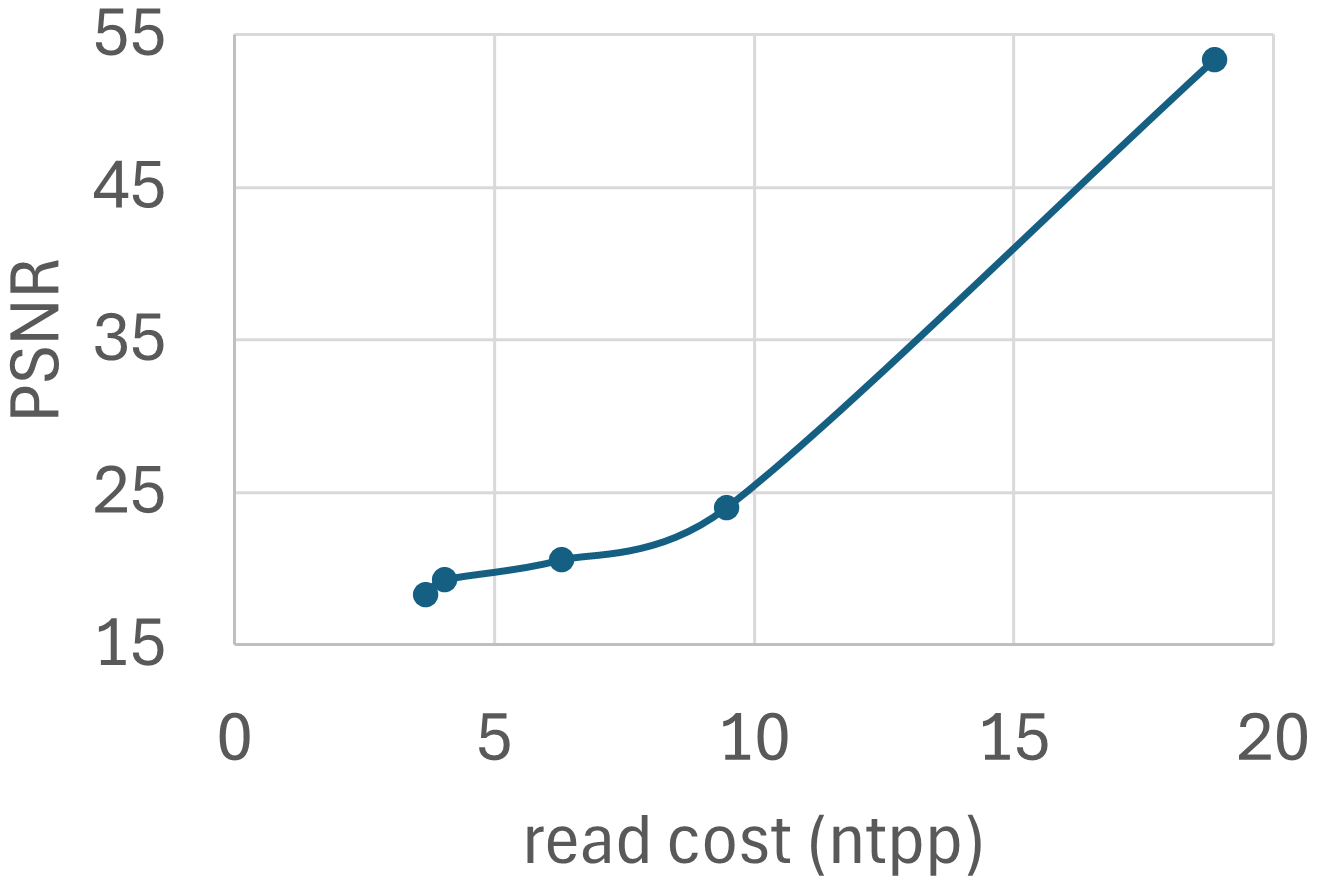}\label{fig:sub_b}}
\vspace*{-0.5\baselineskip}
\end{minipage}
\begin{minipage}{.4\textwidth}
    \subfloat[kodim23 Level 0]{\includegraphics[width=0.5\textwidth]{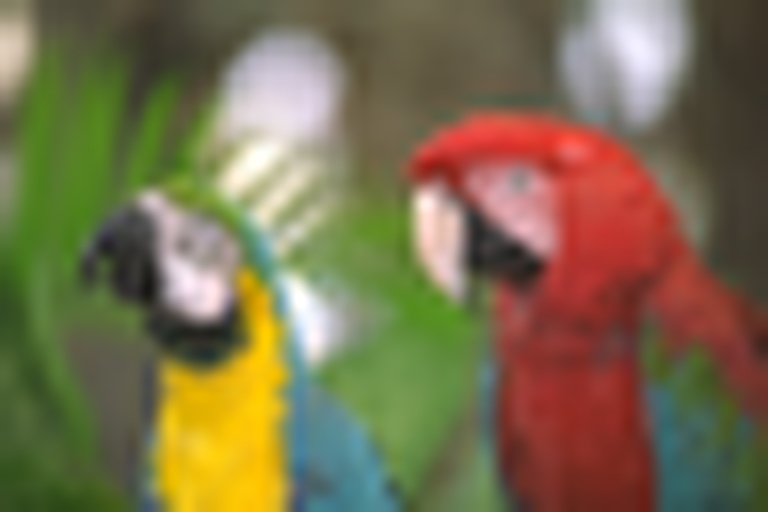}\label{fig:sub_c}}
    \hspace*{1.3cm}
    \subfloat[kodim23 Level 4]{\includegraphics[width=0.5\textwidth]{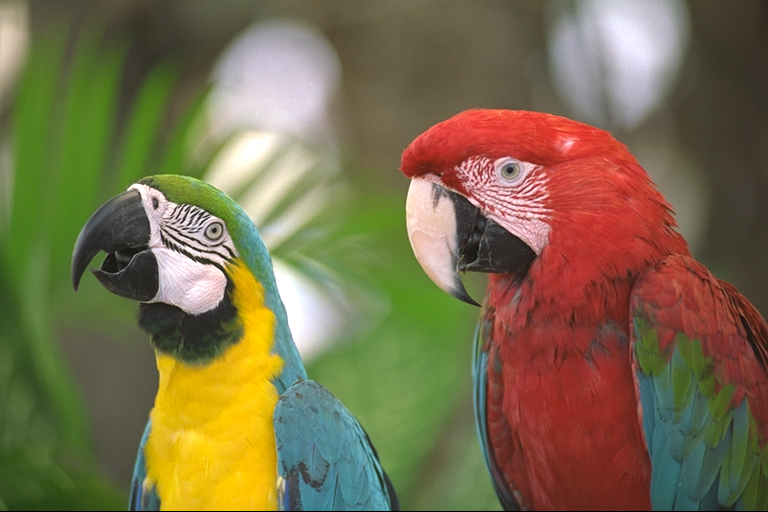}\label{fig:sub_d}}
\end{minipage}
\vspace*{-0.5\baselineskip}
    \caption{ Evolution of the distortion in function of the read-cost through the different resolution layers of the encoded image.}\label{fig:sub}
\vspace*{-1.5\baselineskip}
\end{figure}

Additionally, Figures \ref{fig:sub_a} and \ref{fig:sub_b} depict the evolution of the reconstruction quality (PSNR here) of the image, as a  function of the read cost (which depends on the resolution layers that are selected).
With the reduction factors that we used in the progressive coding parameters, we see that it is possible to decode a degraded version of the image for a fraction of the read-cost necessary to decode the whole picture. For instance, the thumbnail shown in Figure \ref{fig:sub_c} has a PSNR of 18.3dB\footnote{ Additional data can be found here: \\{\url{https://gitlab.eurecom.fr/pic/jp2dnaprogressiveres}}}(to be able to compute the PSNR, the reduced image was resized to the same dimensions as the original image with a bi-cubic interpolation). The last resolution level (Fig. \ref{fig:sub_d}), in contrast, shows a PSNR of 52dB. Further, as can be seen, the thumbnail is of good enough quality to be used as a visual reference in the random access process (Fig. \ref{fig:secondary_primer}). Moreover, the gains we obtain here are orthogonal to the improvements in coding performance and read cost of the different resolution layers that can be achieved by tweaking the encoding options of JPEG2000 (specifically the targeted quality) to further degrade the image. 
Finally, the JPEG DNA VM software parameters can be adjusted differently for each resolution layer, especially the redundancy, so that lower resolution layers are better protected against errors.



\begin{table}[]
    \centering
    \resizebox{8.5cm}{!}{
    \begin{tabular}{cc||c|c|c|c|c|}
        \cline{2-7}
        \multicolumn{1}{c|}{}&Layer & $L_0$ & $L_1$ & $L_2$ & $L_3$ & $L_4$ \\
        \cline{2-7}
        \noalign{\vskip-1\tabcolsep \vskip\arrayrulewidth \vskip-2.2\doublerulesep}
\\ \hline

        \multicolumn{1}{|c||}{\multirow{6}{*}{\rotatebox[origin=c]{90}{Progressive}\hspace*{0.08cm}\rotatebox[origin=c]{90}{Decoding}}}&\# Oligos & 2878 & 8539 & 25288 & 69358 & 151958 \\
        \cline{2-7}
        \multicolumn{1}{|c||}{}&Theoretical $G_{pd}$& \multirow{2}{*}{52.8} & \multirow{2}{*}{17.8} & \multirow{2}{*}{6.01} & \multirow{2}{*}{2.19} & \multirow{2}{*}{1} \\
        \multicolumn{1}{|c||}{}&Read-cost gain &&&&&\\
        \cline{2-7}
        \multicolumn{1}{|c||}{}&Coverage  & 1.67 & 2 & 1.96 & 1.88 & 1.5 \\
        \cline{2-7}
        \multicolumn{1}{|c||}{}&\textbf{Observed $G_{pd}$}& \multirow{2}{*}{\textbf{52.5}} & \multirow{2}{*}{\textbf{15.6}} & \multirow{2}{*}{\textbf{5.13}} & \multirow{2}{*}{\textbf{1.93}} & \multirow{2}{*}{\textbf{1}} \\      
        \multicolumn{1}{|c||}{}&\textbf{Read-cost gain} &&&&&\\
        \hline
        \hline
        \multicolumn{1}{|c||}{\multirow{6}{*}{\rotatebox[origin=c]{90}{Random Access}\hspace*{0.25cm}\rotatebox[origin=c]{90}{\&}\hspace*{0.25cm}\rotatebox[origin=c]{90}{Progressive}\hspace*{0.08cm}\rotatebox[origin=c]{90}{Decoding}}}&\# Oligos & 2878 & 3114 & 3812 & 5648 & 9090\\
        \cline{2-7}
        \multicolumn{1}{|c||}{}&Theoretical $G_{ra}$& \multirow{2}{*}{52.8} & \multirow{2}{*}{48.8} & \multirow{2}{*}{39.9} & \multirow{2}{*}{27.1} & \multirow{2}{*}{17.0}\\      
        \multicolumn{1}{|c||}{}&Read-cost gain &&&&&\\
        \cline{2-7}
        \multicolumn{1}{|c||}{}&Coverage  & 1.67 & 2 & 1.96 & 1.88 & 1.5 \\
        \cline{2-7}
        \multicolumn{1}{|c||}{}&\textbf{Observed $G_{ra}$}& \multirow{2}{*}{\textbf{52.5}} & \multirow{2}{*}{\textbf{47.8}} & \multirow{2}{*}{\textbf{38.4}} & \multirow{2}{*}{\textbf{26.5}} & \multirow{2}{*}{\textbf{17.8}} \\      
        \multicolumn{1}{|c||}{}&\textbf{Read-cost gain} &&&&&\\
        \hline
    \end{tabular}}
\vspace*{-0.5\baselineskip} 
    \caption{ Theoretical and observed average read-cost gains $G_{pd}$ and $G_{ra}$ for each target resolution level, averaged over all the images of the kodak dataset. The level $L_0$ refers to the thumbnail while the level $L_4$ refers to the full image.}
\vspace*{-1.5\baselineskip}
    \label{tab:read_costs}
\end{table}

\vspace*{-0.75\baselineskip}
\section{Conclusion and perspectives}
\vspace*{-0.5\baselineskip}
In this paper, we introduced PIC-DNA--a novel image compression process adapted to DNA data storage that enable new access paths and reduction in read costs. More specifically, PIC-DNA provides a random access solution based on thumbnails that are encoded in specific oligos of the whole oligo pool where data is stored. It additionally provides a trade-off between read-cost and reconstruction quality, by utilizing the progressive functionality of the JPEG2000 image coder. This last functionality can be implemented in DNA storage with any image coders that includes a progressive decoder.

In future work, this software could be used to experiment with other progressive decoders such as JPEG XL.
Additionally, AI-based image coders that include a hierarchical representation of the data, such as \cite{deep_progressive} could be considered as solutions to separate the oligos into sets of oligos of different layers. Videos could also be encoded with a software following the same design. We could store the video, with a thumbnail, into separate video blocks, of small duration, and with different qualities.

\newpage
\bibliographystyle{IEEEbib}
\bibliography{biblio}
\end{document}